# Ultralow Lattice Thermal Conductivity in Complex Structure $Cu_{26}V_2Sn_6Se_{32}$ due to Interaction of Low-Frequency Acoustic-Optical Phonons

Kewal Singh Rana,[1] Debattam Sarkar,[2] Nidhi,[3] Aditya Singh,[1] Chandan Bera,[3]

Kanishka Biswas[2] and Ajay Soni*[1]

[1]School of Physical Sciences, Indian Institute of Technology Mandi, Mandi 175075, Himachal Pradesh, India

[2]New Chemistry Unit, School of Advanced Materials and International Centre of Materials Science, Jawaharlal Nehru Centre for Advanced Scientific Research (JNCASR), Jakkur, Bangalore 560064, India

[3]Institute of Nano Science and Technology, Knowledge City, Sahibzada Ajit Singh Nagar, 140306 Punjab, India

Corresponding Author Email: ajay@iitmandi.ac.in

**Abstract**

Damping of phonon momentum suppresses the lattice thermal conductivity ($\kappa_l$) through low energy acoustic-optical phonon interactions. We studied the thermal transport properties and underlying mechanism of phonon interactions in the large unit cell $Cu_{26}V_2Sn_6Se_{32}$. The large number of atoms in the unit cell results in low acoustic phonon cutoff frequency, flat phonon branches, low frequency Raman active modes, localized rattler-like vibrations and strong crystalline anharmonicity. The crystal structure complexity disrupts the phonon propagation through weak bonded Cu atoms, boson peak and poor phonon velocity. The sulfur at selenium sites ($Cu_{26}V_2Sn_6Se_{30}S_2$) distort the crystal lattice by offering additional scattering mechanism at the anionic sites, thereby increases the power factor and decreases the $\kappa_l$. This strategic manipulation of phonon scattering towards ultra-low $\kappa_l$ not only results in improved thermoelectric performance but also offers insights into the fundamental understanding of heat transport in complex structured, large unit cell compounds.

# 1. Introduction

Complex large unit cell minerals have been studied widely in the fields of thermoelectric (*TE*), thermal barrier coating, and thermal managements in the electronic devices, because of inherent low lattice thermal conductivity ($\kappa_l$).[1-4] Due to abundance, eco-friendly and absence of rare-earth elements (such as Bi, Sb, Pb, Cd and Te), minerals can be future candidates for green energy harvesting technologies.[1-3] In this regard, it is pivotal to demonstrate the fundamental understanding of poor thermal transport of complex crystal structured minerals where the enhanced phonon scattering is an essential requirement.[5-7]

The inherently low $\kappa_l$ in minerals is mainly arising from large number of unit cell atoms,[5,8] robust acoustic-optical phonon interactions,[5,9] soft chemical bonding,[10] rattling-like localized vibrations,[5,10] crystal defects and disorders.[1,11] According to kinetic theory and Boltzmann transport equation: $\kappa_l = \frac{1}{3}(C_v\, v\, l)$, where $C_v$ is the heat capacity, $v$ is the phonon group velocity and $l$ is the phonon mean free path.[7] For $C_v$, the large number of unit cell atoms (*n*) results in (3*n*) vibrational modes strongly hinders the momentum of phonons through strong acoustic-optical phonon interactions.[5-7] The presence of heavy atoms, weak chemical bonding, low elastic modulus altogether lowers the phonons group velocity (*v*).[7] While, the presence of grain boundaries, crystal defects, disorders and nano-structuring lowers the mean free path of phonon (*l*) through the multiple phonon scattering channels.[1,7,11,12]

In *TE*s, generally, the large unit cell minerals are transition metal (Fe, Mn, Zn, Co, Cr, Nb, V, Ni) based ternary and quaternary chalcogenides which shows the greatest impact in the community recently due to inherent low $\kappa_l$ values.[2,4,5] The minerals such as colusites,[1,11,13] tetrahedrites,[14,15] argyrodites,[8] chalcopyrites,[16] sulfide bornite,[17] kuramite,[18] kesterite[19] belong to the family of large unit cell or the complex crystal structure compounds. Colusite is one such mineral where the large number of atoms per unit cell results in extremely low $\kappa_l$ via structural complexity, strong crystal anharmonicity, soft "Cu-S/Se" crystal framework and strong acoustic-optical phonon interaction.[1] The excess negative charge in ions, $(Cu_{26})^+$, $(V_2)^{5+}$, $(Sn_6)^{4+}$ and $(S_{32})^{2-}$ results in electron-deficient character, and shows the heavily degenerate p-type semiconducting behavior [20]. The sulfur sublimation in colusite brings the atomic-scale defects resulting in disordered phase and enhanced phonon scattering. [11,21] Currently, the optimized doping and substitution concentrations at the cationic sites results in saturation of $\kappa_l$ values (~0.4 W/mK).[1,9] Hence, the replacement of sulfur at anionic site with heavy selenium can be an effective route to study the thermal properties due to modified chemical bonding and lattice dynamics. Tailoring at the anion sites with different compositions can also make a better *TE* material through providing a clean pathway for electrical transport and distorted channels for phonon propagation.

Current study explores the poor thermal transport properties in large unit cell synthetic $Cu_{26}V_2Sn_6Se_{32}$ (CVS-Se$_{32}$) and sulfur doped $Cu_{26}V_2Sn_6Se_{30}S_2$ (CVS-Se$_{30}$S$_2$) materials. In CVS-Se$_{30}$S$_2$, the poor average sound velocity large and Gruneisen parameter reveal the presence of crystal anharmonicity. The large number of primitive unit cell atoms strongly hinders the efficiency of heat carrying acoustic phonons through low acoustic phonon cut-off frequency. The low frequency optical phonons strongly interact with acoustic phonons and boosts the scattering mechanism. The presence of boson peak, existence of localized Einstein vibrational modes, suggest the presence of disorder and triggers the multi-phonon scattering processes. Among all the cations, the thermal and vibrational properties are mainly delivered by the weakly bonded Cu ions. The experimental studies are supported by the first principle calculations through the electronic and phononic dispersion curves and density of states.

## 2. Experimental and Computational Details

The polycrystalline samples $Cu_{26}V_2Sn_6Se_{32}$ and $Cu_{26}V_2Sn_6Se_{30}S_2$ were synthesized through the solid-state melting route. High purity (~ 99.99%) raw elements (Cu, V, Sn, Se and S) were weighed in stoichiometric amount and placed in clean quartz ampoules. The flame-sealed quartz ampoules (~$10^{-5}$ mbar) were slowly heated up to ~ 1323 K and kept for 48 hours and cooled down to room temperature.[9] The obtained ingots were grounded into fine powder and subjected to spark plasma sintering (SPS) process to consolidate the sample for a period of ~10 minutes at ~773 K under a uniaxial pressure of ~30 MPa and vacuum ~$10^{-3}$ mbar. The density ($d_m$) of obtained SPS-processed coins were found to be ~ 5.50 g cm$^{-3}$ for both materials.

The phase purity and crystal structure were carried out through X-Ray diffraction (XRD) pattern with rotating anode Rigaku Smart lab diffractometer and Cu-K$_\alpha$ radiation (wavelength ~ 1.5406 Å). The field emission scanning electron microscopy (FE-SEM) images and elemental mapping were collected through JFEI, USA, Nova Nano SEM-450. The Raman spectroscopic measurements were carried out through Horiba Jobin-Yvon LabRAM HR evolution, 532 nm excitation laser and 1800 grooves/mm grating. The ultra-low frequency filters were used to identify the low frequency modes. The temperature dependent (300 - 773K) electrical conductivity ($\sigma$) and Seebeck coefficient ($S$) measurements were performed using ZEM-3 (ULVAC-RIKO) on a bar-shaped pellets under He atmosphere. The charge carrier concentration and low temperature $C_p$ (specific heat) experiments were carried out through physical property measurement system (PPMS, Quantum Design). The total thermal conductivity ($\kappa_{total}$) was estimated using, $\kappa_{total} = DC_p d_m$, where $D$ is the diffusivity. Further, the laser flash analysis (LFA) was used to measure $D$ under a nitrogen atmosphere through Netzsch LFA-457, and $C_p$ was estimated through the Dulong-petit approximation. The longitudinal ($v_l$) and transverse ($v_t$) sound velocities were measured on the disc-shaped SPS-processed samples through an Epoch 650 Ultrasonic Flaw Detector (Olympus) having the transducer frequency 5 MHz.

Theoretical calculations were done inside the framework of Density Functional Theory (DFT), the electronic structure was investigated within the Vienna Ab initio Simulation Package (VASP).[22,23] The kinetic cutoff energy of 420 eV was used to expand the wave-function in basis set of plane waves in Projected Augmented Wave (PAW) method, and Perdew-Burke-Ernzerhof (PBE) exchange correlation was applied. [22,24,25] In combination with above a k-mesh of 4*4*4 was sampled in the reciprocal space. The convergence criterion of total energy was set to $10^{-8}$ eV and full optimization of structures was done until the force on each atom is 0.05 eV Å$^{-1}$. The crystal structure was relaxed to cubic crystal structure and space group symmetry $P\bar{4}3n$ with lattice parameters a ~11.3136 Å and ~ 11.3064 Å for CVS-Se$_{32}$ and CVS-Se$_{30}$S$_2$, respectively. Furthermore, Phonopy software was used to calculate the phonon dispersion curve and a supercell of size 1*1*1 and 1*1*1 k-points was considered for both the structures.[26]

## 3. Result and Discussion

**(i) X-ray Diffraction**

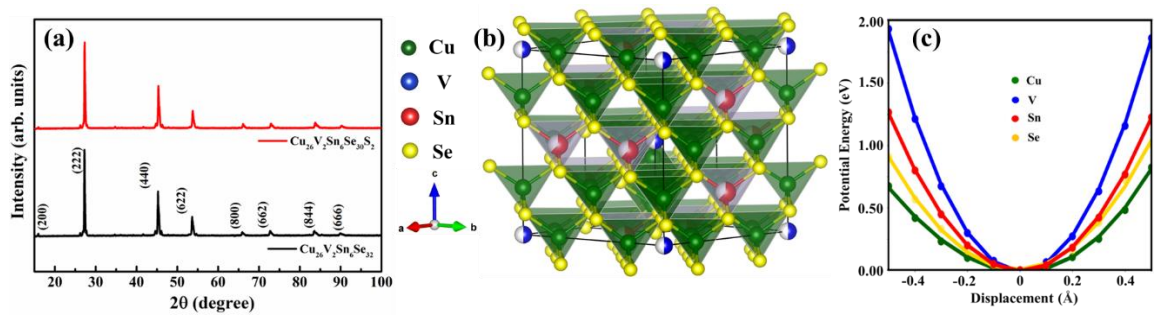

Figure 1. Powder XRD patterns of (a) CVS-Se$_{32}$ and CVS-Se$_{30}$S$_2$, (b) the cubic polyhedral structure of CVS-Se$_{32}$, and (c) potential energy vs relative movement of all elements representing the relatively rigid nature of V and Sn cations with respect to soft bonded Cu cations.

Figure 1 (a) shows the XRD patterns of finely ground powder samples of CVS-Se$_{32}$ and CVS-Se$_{30}$S$_2$. Most of reflections are matching with earlier results of Colusites having space group $P\bar{4}3n$ and cubic crystal structure.[1] However, few reflections with very weak intensity (~26.2, 34.7 and 41.6°) refers the presence of secondary phases (Cu and Se phases). The Rietveld refined XRD pattern is shown in figure S1, having refined lattice parameters ($a = b = c$) ~11.3155 Å and unit cell volume (V$_c$) ~1448.84 Å$^3$ for cubic CVS-Se$_{32}$. Further, the right shift of (222) diffraction peak of CVS-Se$_{30}$S$_2$ with respect to CVS-Se$_{32}$, represents the lattice contraction ($a$ ~11.2881 Å and V$_c$ ~1438.34 Å$^3$) due to lower radii sulfur in selenium sites. In comparison with earlier study of Cu$_{26}$V$_2$Sn$_6$S$_{32}$ Colusite, the larger radii Se atom enhances V$_c$ in Cu$_{26}$V$_2$Sn$_6$Se$_{32}$ by a factor of ~15%, signifying the enhanced unit cell dimensions.[27] The cubic crystal structure of CVS-Se$_{32}$ is shown in figure 1 (b) with five cationic sites, Cu atom occupies 12$f$, 8$e$ and 6$d$, V occupies 2$a$ and Sn occupies 6$c$ atom; and Se atom occupies two anionic sites at 24$i$ and 8$e$. All cations are tetrahedrally coordinated with Se anion and construct a large three dimensional (Cu/V/Sn)-Se$_4$ network structure having 66 atoms in the primitive unit cell.[1] The microstructural FE-SEM images and elemental color mapping (Figure S2) on clean and polished surface verifies the homogeneity and the uniform distribution of all the constituent elements. Furthermore, among all the cations, the shallow potential well of Cu atoms compared to Sn and V atoms shows the presence of soft or weak chemical bonding (Figure in 1 (c)). A small perturbation or thermal excitation is sufficient to disturb the Cu atoms from their equilibrium positions. The deeper potential energy well for Sn and V atoms shows the presence of rigidity in the crystal lattice. In this regard, the effect of Cu disturbance can play a vital role towards the thermal and vibrational properties of these materials.

**(ii)    Raman measurements**

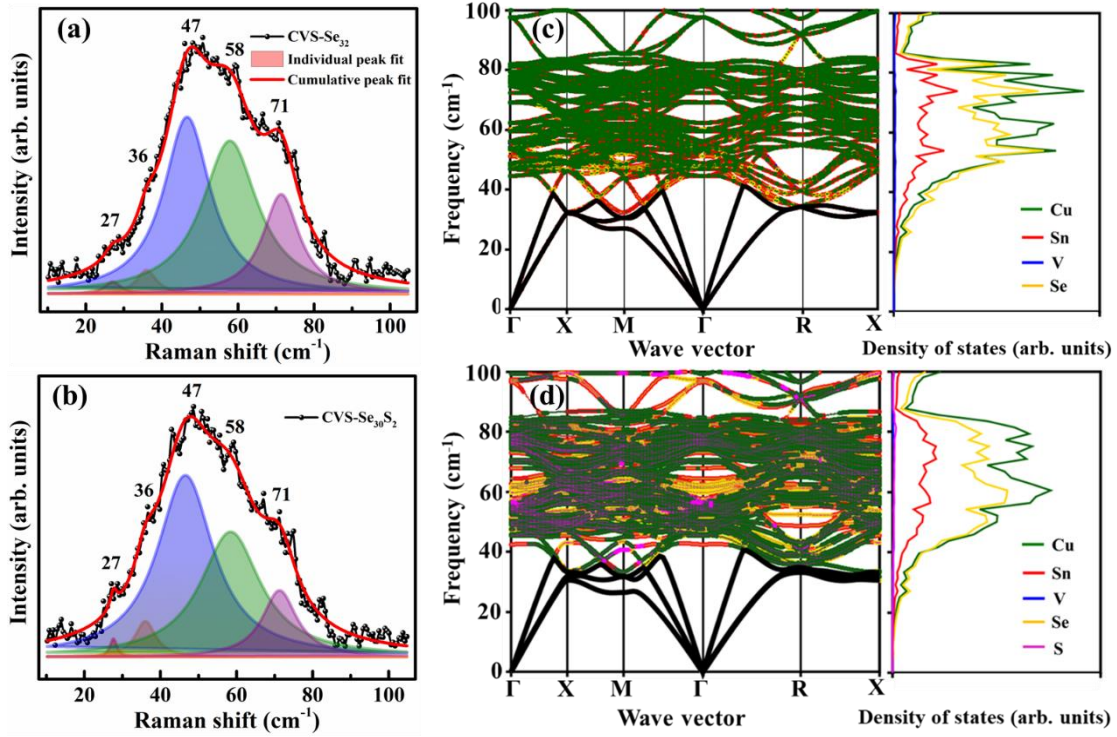

Figure 2. Room temperature (a, b) Raman spectra. (c, d) Phonon dispersion curve and corresponding atom projected density of states for CVS-Se$_{32}$ and CVS-Se$_{30}$S$_2$, respectively.

Figure 2 (a) and (b) represents the Raman spectra of CVS-Se$_{32}$ and CVS-Se$_{30}$S$_2$ in the low-frequency regimes where the acoustic-optical phonon interactions play significant role towards thermal transport. Structurally, the large unit cell compounds with 66 atoms in the unit cell ($n$) results in maximum 195 (3$n$-3) optical modes with $A_1$, $E$, $T_1$ and $T_2$ possible Raman mode symmetries. We have observed five low-frequency Raman active modes (below 100 cm$^{-1}$) for both materials, positioned at ~ 27, 36, 47 ($T_1$), 58 ($T_1$) and 71 cm$^{-1}$ ($A_2$). The phonon dispersion curve (Figure 2 (c)) shows the flat and compressed optical modes from ~ 40 to ~ 80 cm$^{-1}$, mainly associated with the vibration of soft bonded Cu atoms. The Cu cations are strongly vibrating relatively among all the cations as evident from the atom projected density of state calculations (Figure 2 (c)). The eigen vector visualization of Raman active modes are shown in figure S3. At $X$, $M$ and $R$ symmetry points, the lowest possible optical phonon modes are observed (Figure 2 (c)) and the eigen vector visualization are shown in figure S4. The interaction of these modes with heat carrying acoustic branches changes the phonon's momentum. The sulfur incorporation provides the disorder in the anionic sites of crystal lattice which reduces the overall stiffness of the lattice and decreases the effective mass through the mass discrepancy. Furthermore, the unattributed peaks at ~ 27 and 36 cm$^{-1}$ may be related to the crystal structural distortions. Overall, a compress and almost flat phonon dispersion curve is observed (Figure 2 (d)), which is clearly signifying the poorer sound propagation and strong crystal anharmonicity in CVS-Se$_{30}$S$_2$ than CVS-Se$_{32}$. For both compounds, the atomic density of states calculations suggesting that the vibrational properties are highly dominated by the softly bonded CuSe$_4$ tetrahedra. The acoustic branches are lies below 40 cm$^{-1}$, further signifying the poor thermal properties in these complex structured materials, due to their low acoustic phonon cut-off frequency. We have also calculated the phonon lifetime ($\tau_i$) of the active modes through the obtained full-width half-maxima (FWHM's) via, $\tau_i = \frac{1}{2\pi\ FWHM_i}$, and obtained $\tau_i \leq 1$ ps.[9] The short $\tau_i$ of Raman active modes suggests the faster phonons scattering processes.

### (iii) Electrical transport

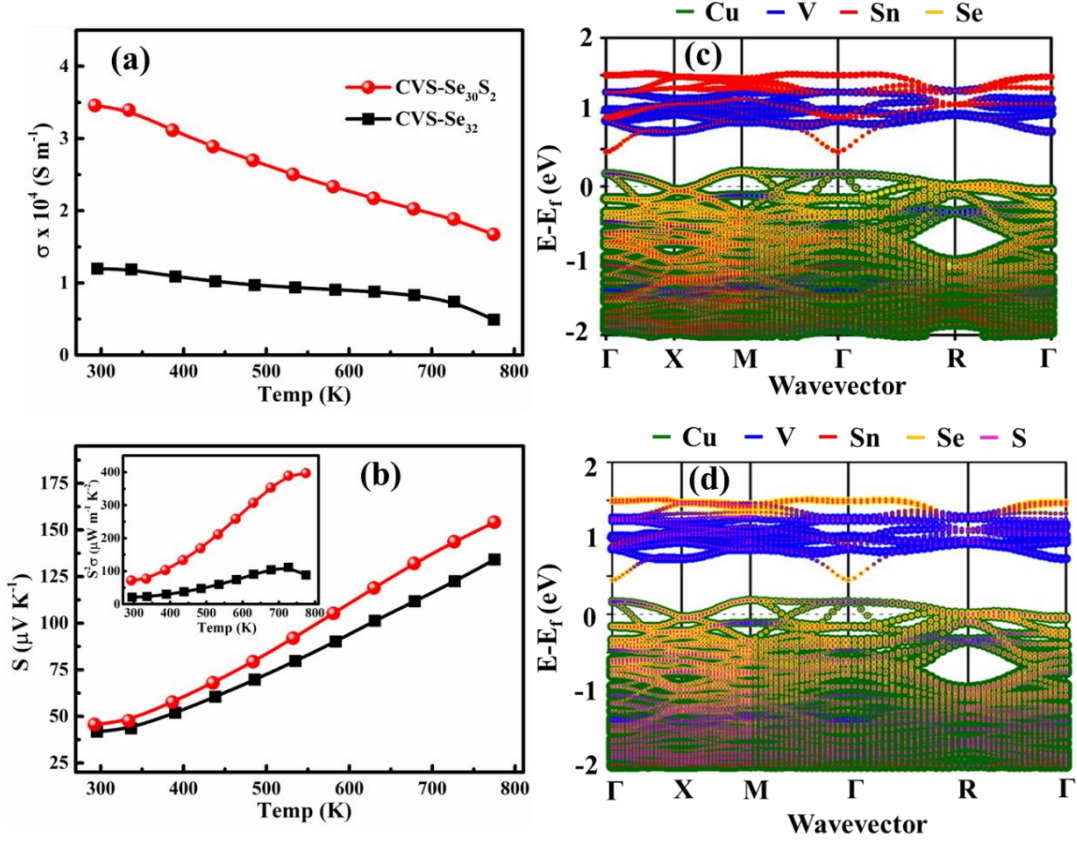

Figure 3. Temperature dependent (a) electrical conductivity, $\sigma$, (b) Seebeck coefficient, $S$, inset, represents the power factor, $S^2\sigma$, and (c, d) electronic dispersion curve for CVS-Se$_{32}$ and CVS-Se$_{30}$S$_2$, respectively.

Temperature dependence of electrical conductivity ($\sigma$) decreases with the increase in temperature suggesting the degenerate semiconductor behavior (Figure 3 (a)). The ordered CVS-Se$_{32}$ compound has the lowest $\sigma$, whereas, the sulfur doped CVS-Se$_{30}$S$_2$ has increased $\sigma$. The positive Seebeck coefficient ($S$) shows that holes are the majority charge carriers (Figure 3 (b)) and lies in the range of ~ 50 - 150 μV K$^{-1}$ for 300 - 773 K. The electrical transport properties are also verified through the theoretical calculations. The electronic dispersion curve calculations further suggesting similar results as observed experimentally. The extended valance band states above the Fermi level ($E_f$) ((Figure 3 (c) and (d))) implies holes are the major charge carriers. Further, the room temperature carrier concentration is ($n_H$) ~ 3.7 x 10$^{20}$ cm$^{-3}$ and ~ 8.6 x 10$^{20}$ cm$^{-3}$ for CVS-Se$_{32}$ and CVS-Se$_{30}$S$_2$, respectively. Here, by introduction of sulfur dopants the enhances the carrier concentration and scattering mechanism without affecting significantly the density of states near $E_f$. The inset figure 3 (c) shows high power factor ($PF = \sigma S^2$) ~ 400 µW m$^{-1}$ K$^{-2}$ in CVS-Se$_{30}$S$_2$, whereas the ordered CVS-Se$_{32}$ compound shows low $PF$ ~90 µW m$^{-1}$ K$^{-2}$ at ~773 K.

**(iv)    Thermal Transport Measurements**

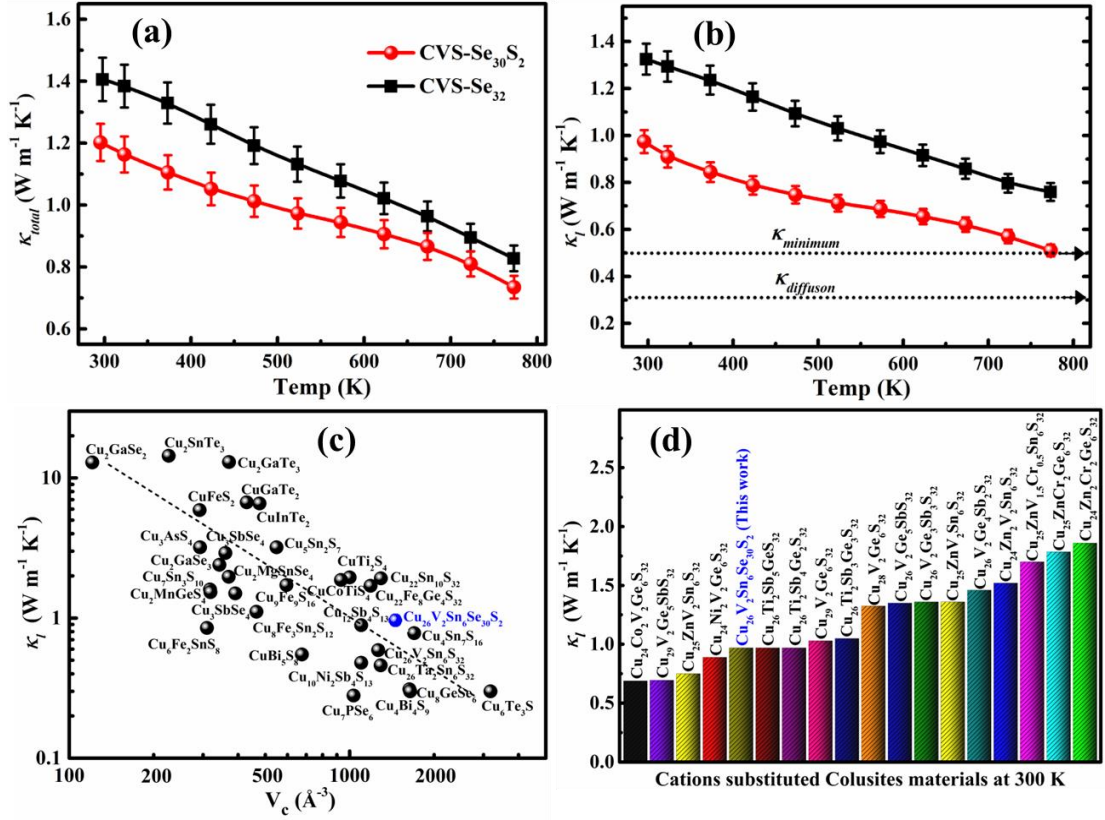

Figure 4. Temperature dependent (a) total thermal conductivity ($\kappa_{total}$), and (b) lattice thermal conductivity, ($\kappa_l$) for CVS-Se$_{32}$ and CVS-Se$_{30}$S$_2$. (c) The variations of $\kappa_l$ vs unit cell volume ($V_c$) in Cu-based complex crystal structure chalcogenides [12,15,27-38] and, (d) comparison of room temperature $\kappa_l$ for various cations and anions substituted colusites compounds having multiple phonon scattering channels.[39-44]

Figure 4 (a) shows the low and decreasing $\kappa_{total}$ behaviour for CVS-Se$_{32}$, and CVS-Se$_{30}$S$_2$ with rise in the temperature. The low $\kappa_{total}$ in these compounds is due to the large number of atoms in the primitive unit cell ($N$), where $N$ shows inverse dependence with $\kappa_{total}$, hence, lower the heat transportation, significantly.[8,12] At higher temperature, the amplitude of phonons vibration increases leading to scattering which further lowers the $\kappa_{total}$. The sulfur substitution in the regular crystal framework of CVS-Se$_{32}$ provides the atomic scale disorder and largely disrupts the heat-carrying phonon paths. Furthermore, the $\kappa_l$ is calculated by subtracting the electronic thermal conductivity ($\kappa_e$) from $\kappa_{total}$ as, $\kappa_l = \kappa_{total} - \kappa_e$, where the $\kappa_e$ (Figure S8) is estimated via the Wiedemann-Franz law, $\kappa_e = \sigma LT$, here $L = [1.5 + \exp(-|S|/116)]$ is the Lorenz number and $|S|$ is the absolute Seebeck coefficient.[45] The obtained $\kappa_l$ shows a decrease trend with increase in temperature (Figure 4 (b)) due to anharmonic *Umklapp* processes ($T^{-1}$ dependence).[46] At high temperature, the phonon-phonon scattering becomes dominant, creating additional phonon scattering channels which further lowers to glass-like $\kappa_l$ value ~ 0.51 W m$^{-1}$ K$^{-1}$ (at 773 K) in CVS-Se$_{30}$S$_2$. The experimental observed $\kappa_l$ is just above the minimum theoretical thermal conductivity ($\kappa_{minimum}$) which is calculated using the Cahill's Model for pristine CVS-Se$_{32}$.[47] The diffusive-mediated thermal conductivity ($\kappa_{diffuson}$) is the lowest possible experimental thermal conductivity for bulk disordered materials related to the average energy of vibrational density of states of a materials and the number density of atoms.[48] The enhanced disorder with S doping in CVS-Se$_{30}$S$_2$ further lowers the $\kappa_l$ of CVS-Se$_{32}$ to the $\kappa_{minimum}$ level. Furthermore, in general, with increasing the unit cell volume, the $\kappa_l$ of the material decreases.[5,7,12] Figure 2 (c) shows the inverse relation between $\kappa_l$ and unit cell volume (V$_c$) for various complex structured large unit cell materials. Moreover, figure 2 (d) shows the comparison of room temperature $\kappa_l$ of CVS-Se$_{30}$S$_2$ and various cation substituted and doped Colusites materials. The significant low $\kappa_l$ in CVS-Se$_{30}$S$_2$ represents the potential towards the thermal applications.

**(v)    Heat Capacity Analysis**

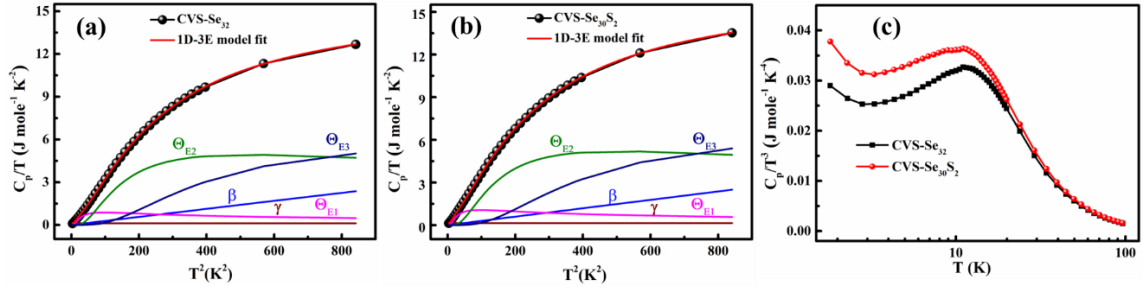

Figure 5. Temperature dependent $C_p$ for (a) CVS-Se$_{32}$, (b) CVS-Se$_{30}$S$_2$. The data is best fitted using 1D-3E model due to localized vibrational modes at low-energies, and (c) boson peak representation via $C_p/T^3$ vs $T$ plot, shows the presence of disorder and relatively high vibrational density of states in CVS-Se$_{30}$S$_2$.

The detailed analysis about the localised vibrations, disorders, crystal anharmonicity, and acoustic-optical phonon interaction, we have studied and explored the low temperature $C_p$ properties (Figure 5 (a) and (b)). Using single Debye model fitting of $C_p/T$ vs $T^2$, the negative value of electronic heat capacity ($\gamma$) term signify that the $C_p$ analysis can't be examined from single Debye model alone.[49] Therefore, we try to fit the temperature dependence $C_p$ data through "Debye and additional localized Einstein modes", as represented by following equation:

$$\frac{C_p}{T} = \gamma + \beta T^2 + \sum_{n=1}^{3}\left(A_n(\Theta_{En})^2 T^{2(-\frac{3}{2})}\frac{e^{\frac{\Theta_{En}}{T}}}{\left(e^{\frac{\Theta_{En}}{T}} - 1\right)^2}\right)$$

where first, second and third term represents the electronic (Sommerfeld), lattice (Debye) and multiple localized oscillators (Einstein) contributions of heat capacity, respectively. Here $A_n$ is the prefactor and $\Theta_{En}$ is the Einstein temperature's for the $n^{\text{th}}$ Einstein oscillators.[50] For CVS-Se$_{32}$ the three characteristic localized Einstein modes positioned at ~25 (17), 60 (42) and 100 K (70 cm$^{-1}$), interacts with acoustic and optical modes in the low frequency regions. For CVS-Se$_{30}$S$_2$, the similar fitting parameters are shown in the Table S1 (supplemental material). Hence, the deviation in the Debye $T^3$ law (non-Debye behaviour) can be understand by the presence of these excess vibrational density of states. Further, figure 5 (c) shows a peak at low temperature, known as "Boson peak", a characteristic peak for materials having (high) disorder. The significant high $C_p/T^3$ shift of boson peak with higher magnitude of peak value suggests the presence of larger vibrational density of states and more glassy nature in CVS-Se$_{30}$S$_2$.[51]

**(vi) Sound Velocity Measurements and Crystal Anharmonicity**

Table 1. The longitudinal ($v_l$), transverse ($v_t$), and average sound velocities ($v_m$); minimum ($\kappa_{minimum}$) and diffuson ($\kappa_{diffuson}$) thermal conductivities; Poisson ration ($v_p$), Debye temperature ($\Theta_D$), Grüneisen parameters ($\gamma_G$), bulk modulus (B) and shear (G) modulus of complex structured CVS-Se$_{32}$ and CVS-Se$_{30}$S$_2$ materials.

|  | **CVS-Se$_{32}$** | **CVS-Se$_{30}$S$_2$** |
|---|---|---|
| Longitudinal sound velocity, $v_l$ (m s$^{-1}$) | ~3423 | ~3150 |
| Transverse sound velocity, $v_t$ (m s$^{-1}$) | ~1786 | ~1450 |
| Average sound velocity, $v_m$ (m s$^{-1}$) | ~1998 | ~1634 |
| Minimum thermal conductivity, $\kappa_{minimum}$ (W m$^{-1}$ K$^{-1}$) | ~0.50 | ~0.43 |
| Diffuson thermal conductivity, $\kappa_{diffuson}$ (W m$^{-1}$ K$^{-1}$) | ~0.31 | ~0.27 |
| Poisson ratio, $v_p$ | ~0.31 | ~0.37 |
| Debye temperature $\Theta_D$, (K) | ~212 | ~174 |
| Grüneisen parameters, $\gamma_G$ | ~1.86 | ~2.27 |
| Bulk modulus, $B$ (MPa) | ~41.0 | ~38.9 |
| Shear modulus, G (MPa) | ~17.5 | ~11.5 |

The relatively soft vibrations of Cu atoms and rattling-like localized vibrations at low frequency regime can results in the soft elastic lattice and poor sound velocity. To study in detail, we have estimated the longitudinal ($v_l$), transverse ($v_t$) and average ($v_m$) sound velocities as listed in Table 1. The observed $v_m$ ~1998 m sec$^{-1}$ (CVS-Se$_{32}$) and ~1634 m sec$^{-1}$ (CVS-Se$_{30}$S$_2$) again signify the poor thermal transport properties. The experimentally values are well supported by the theoretical results, as evident from the flat vibrational branches (less slope in dispersion curve) and the strong interactions among them. In the large unit cell compounds, the large real space dimensions (lattice parameters) leads in the shrinkage of first Brillouin zone dimensions (in reciprocal space), which results in folded back of high-frequency vibrational modes as compressed and flat optical modes.[6] Hence, the flat optical branches in the dispersion curves carries negligible heat energy, so most of heat energy is carried by the acoustic branches. The poor sound propagation and strong acoustic-optical interaction also lowers the crystal's highest normal mode of vibrations, so the Debye temperature ($\Theta_D$). The obtained $\Theta_D$ is lowest than many Cu-based complex structured materials (Figure S9). We have also quantified the Grüneisen parameter ($\gamma_G$), which informs about the presence of crystal anharmonicity and characterise the connection between the crystal volume, phonon frequency and $\kappa_l$. The strong crystal anharmonicity (large $\gamma_G$) causes large damping effects, enhances the scattering processes and ultimately lowers the $\kappa_l$. The obtained $\gamma_G$ are ~ 1.86 (CVS-Se$_{32}$) and ~2.27 (CVS-Se$_{30}$S$_2$), which is significant higher than many state-of-the-art *TE* materials like Bi$_2$Te$_3$, Sb$_2$Te$_3$ PbTe, PbSe and PbS.[52] The disorder by the sulfur atom in the crystal lattice of regular arranged Se sites, soften the Cu-Se/S$_4$ tetrahedra, so enhances the crystal anharmonicity in CVS-Se$_{30}$S$_2$ .[53] The low-acoustic phonon cut-off frequency (~ 40 cm$^{-1}$ for CVS-Se$_{32}$, figure 3(c)) also results in poor elastic properties (bulk and shear modulus) due to soft chemical bonding by Cu atoms. Figure S10 - S11 shows the comparison of low bulk and shear modulus for CVS-Se$_{32}$ and CVS-Se$_{30}$S$_2$ than other complex structured compounds, which makes them a potential candidate for poor elastic and thermal transport applications. Regarding the *TE* figure of merit ($ZT = \frac{\sigma S^2}{\kappa_{total}} T$) perspective, ~ 6% sulfur doped CVS-Se$_{30}$S$_2$ has *ZT* ~ 0.42 at 773 K (Figure S12), which is almost five-time higher than the CVS-Se$_{32}$.

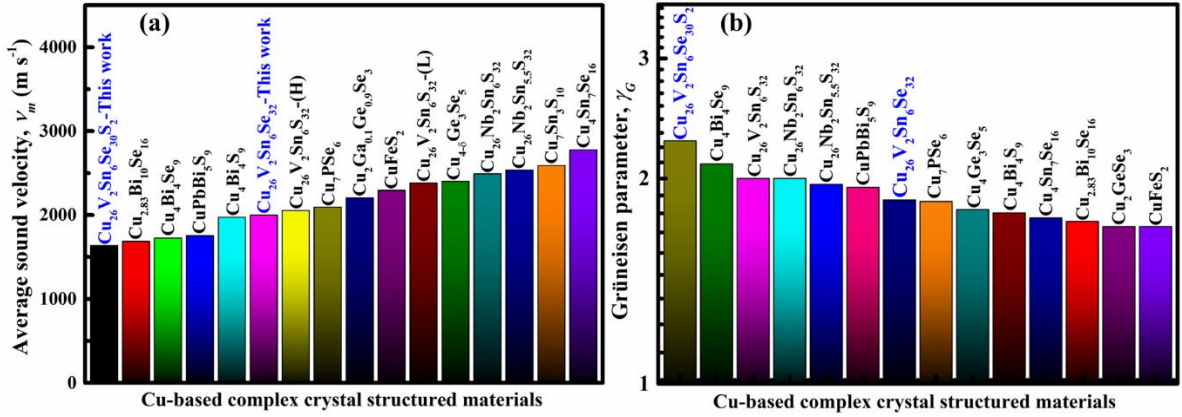

Figure 6. Comparison of (a) average sound velocity ($v_m$) and (b) Grüneisen parameters, ($\gamma_G$) of eco-freiendly Cu-based complex structured sulfides and selenides.[1,31,37,54-59] The $v_m$ for all these materials is calculated from the reported $v_l$ and be $v_t$ values, by using the formula, $v_m = \left(\frac{3}{v_l^{-3} + 2v_t^{-3}}\right)^{1/3}$, where $v_l$ and be $v_t$ be the longitudinal and transverse sound velocities, respectively. Whereas, $\gamma_G$ for all these compounds is calculated through the formula, $\gamma_G = \frac{3}{2}\left(\frac{1+v_p}{2-3v_p}\right)$, here, $v_p$ is the Poisson ratio of the materials.[60]

Overall, the weakly bonded Cu atoms in the crystal framework results in poor sound propagation (low $v_l$ and $v_t$), soft elastic lattice properties, which altogether increases the phonon damping and strongly hinders the phonons transportations. The presence of disorders at the anionic sites of CVS-Se$_{30}$S$_2$ leads to low $v_m$ and high $\gamma_G$ values (Figure 6 (a) and (b)) make poor thermal conductive material than many Cu-based large unit cells, complex structured, and ecofriendly sulfides and selenides.

## 4. Summary

In summary, the large number of atoms in the unit cell can brings the different energy scaled acoustic and optical phonon branches together. The strong interaction between the branches hinders the heat-carrying phonon's momentum strongly by virtue of which the $\kappa_l$ of the materials can be tuned. The poor sound propagation and large Gruneisen parameter, weakly bonded Cu atoms altogether reveals about the strong acoustic-optical phonon interactions and poor thermal properties in the complex structured large unit cell materials. The low acoustic phonons cut-off frequency, flat dispersion curves further signify that the most of the heat is carried by the acoustic branches. The presence of several low-frequency optical modes is originated from the soft bonded Cu atom vibrations. Further, the presence of structural disorder leads deviation in the Debye $T^3$ law through the presence of localized Einstein vibrations and boson peak. The tailoring of sulfur atom in Selenium sites affects both electrical and thermal transport properties, significantly. Overall, due to ultralow $\kappa_l \sim 0.5$ W m$^{-1}$ K$^{-1}$ in CVS-Se$_{30}$S$_2$, the copper based complex structured material can be considered as a potential candidate for future energy harvesting applications.

**ORCID:**

Kewal Singh Rana: 0000-0003-0345-6225

Chandan Bera: 0000-0002-5226-4062

Kanishka Biswas: 0000-0001-9119-2455

Ajay Soni: 0000-0002-8926-0225

**Notes:** The authors declare no competing financial interest.

**Acknowledgements:** A.S. acknowledges IIT Mandi for research facilities.